\documentclass[
aps,
prl,
twoside,
twocolumn,
10pt,
floatfix,
showpacs,
citeautoscript,
]{revtex4-2}

\usepackage{amsmath,amssymb}
\usepackage{comment}
\usepackage{glossaries}
\usepackage{graphicx}
\usepackage[version=4]{mhchem}
\usepackage{siunitx}
\usepackage{tabularx}
\usepackage{hyperref}
\usepackage{todonotes} 

\newcommand \Title{
Thermal Stabilization of Defect Charge States\texorpdfstring{\\}{} and Finite-Temperature Charge Transition Levels
}

\hypersetup{
pdfauthor={T. Hainer, E. Berger, E. Berger, O. Hildeberg, P. Erhart, and J. Wiktor},
pdftitle={\Title},
pdffitwindow=false,
pdfstartview={FitH},
pdfnewwindow=true,
colorlinks=true,
linkcolor=black,
citecolor=black,
filecolor=black,
urlcolor=black,
}

\graphicspath{{./figures/}}

\setacronymstyle{long-short}
\newacronym{cbm}{CBM}{conduction band minimum}
\newacronym{ctl}{CTL}{charge transition level}
\newacronym{dft}{DFT}{density functional theory}
\newacronym{dos}{DOS}{density of states}
\newacronym{fnv}{FNV}{Freysoldt, Neugebauer and Van de Walle}
\newacronym{mlip}{MLIP}{machine-learned potential}
\newacronym{md}{MD}{molecular dynamics}
\newacronym{nep}{NEP}{neuroevolution potential}
\newacronym{pes}{PES}{potential energy surface}
\newacronym{rmse}{RMSE}{root mean square  error}
\newacronym{rs}{RS}{reversible scaling}
\newacronym{ti}{TI}{thermodynamic integration}
\newacronym{vbm}{VBM}{valence band maximum}
\newacronym{xc}{XC}{exchange-correlation}

\DeclareSIUnit\angstrom{\text{Å}}
\DeclareSIUnit\atom{\text{atom}}
\DeclareSIUnit\step{\text{step}}

\makeatletter
\let\oldtheequation\theequation
\renewcommand\tagform@[1]{\maketag@@@{\ignorespaces#1\unskip\@@italiccorr}}
\renewcommand\theequation{(\oldtheequation)}
\makeatother

\newcommand{\addchalmers}{Department of Physics, Chalmers University of Technology, SE-41296, Gothenburg, Sweden}

\begin{document}

\title{\Title}

\author{Tobias Hainer}
\author{Ethan Berger}
\author{Esmée Berger}
\author{Olof Hildeberg}
\author{Paul Erhart}
\author{Julia Wiktor}
\email{julia.wiktor@chalmers.se}
\affiliation{\addchalmers}

\begin{abstract}
Point defects introduce localized electronic states that critically affect carrier trapping, recombination, and transport in functional materials. 
The associated \glspl{ctl} can depend on temperature, requiring accurate treatment of vibrational and electronic free-energy contributions.
In this work, we use machine-learned interatomic potentials to efficiently compute temperature-dependent \glspl{ctl} for vacancies in \ce{MgO}, \ce{LiF}, and \ce{CsSnBr3}. 
Using thermodynamic integration, we quantify free-energy differences between charge states and calculate the vibrational entropy contributions at finite temperatures. 
We find that \glspl{ctl} shift with temperature in \ce{MgO}, \ce{LiF} and \ce{CsSnBr3} from both entropy and electronic contributions. 
Notably, in \ce{CsSnBr3} a neutral charge state becomes thermodynamically stable above \qty{60}{\kelvin}, introducing a temperature-dependent Fermi-level window absent at \qty{0}{\kelvin}.
We show that the widely used static, zero-kelvin defect formalism can miss both quantitative \gls{ctl} shifts and the qualitative emergence of new stable charge states.
\end{abstract}

\maketitle

\clearpage

\glsresetall{}


Defects play a crucial role in determining the physical properties of materials, influencing mechanical, electronic, and thermal behavior \cite{LUBOMIRSKY20061639, RevModPhys.50.797, PhysRevB.82.085436, doi:10.1021/acs.chemmater.1c02345, QU20113841}.
In particular, in functional applications such as photovoltaics they can both critically enhance and degrade the performance of solar energy devices \cite{pantelides_electronic_1978, ball_defects_2016, park_point_2018}. 
Under realistic operating conditions, defects are unavoidable, making it essential to understand their impact on material behavior. 
In non-metals, charged defects can form and interact strongly with their environment through long-range Coulomb forces, influencing electronic and ionic processes. 
A key property of such defects is the \gls{ctl}, which determines the thermodynamically favorable charge state for a given Fermi level. 
These \glspl{ctl} are strongly influenced by the local chemical environment and therefore can be sensitive to temperature \cite{10.1063/1.4975033}.
Accurate predictions of \glspl{ctl} are essential for understanding defect-related properties, underscoring the need for detailed studies of their behavior including finite-temperature effects.

These defect-related properties have mainly been modeled using \gls{dft}, an accurate but computationally expensive method that becomes prohibitive for larger systems and/or is limiting when thermal effects are important.
Many studies have mapped out \glspl{ctl} at zero temperature for a wide range of materials \cite{doi:10.1021/acsenergylett.8b01212, doi:10.1021/acs.jpclett.5b00199, PhysRevMaterials.5.035405, LinLinErh18, RevModPhys.86.253, https://doi.org/10.1002/pssb.201046195}.
While this allows a static characterization of defects, capturing behavior at finite temperatures remains a significant challenge.
At finite temperatures, contributions from spin, electronic, vibrational and orientational entropy play a role in the free energy formulations \cite{mishin_calculations_2001, walsh_imperfections_2023}.
Among these, vibrational entropy typically has the largest impact and must be accounted for to accurately predict defect formation energies at finite temperatures.
Incorporating this temperature dependence is therefore essential for reliable predictions above zero kelvin.

Finite-temperature effects have been explored within the (quasi-)harmonic approximation but also using molecular dynamics approaches \cite{CohEggRap19, walsh_imperfections_2023, LinOstWik25, mosquera2025dynamic}, as well as through analyses of defect-induced volume changes and band-edge fluctuations \cite{PhysRevB.105.115201}.
These studies show that temperature induces only modest variations in defect formation free energies.
However, a comprehensive understanding of the temperature dependence of \glspl{ctl} is still emerging.

Here, we introduce a general finite-temperature framework for defect \glspl{ctl} by combining \gls{dft}-trained \gls{nep} models with \gls{ti} to quantify the temperature dependence of \glspl{ctl}, as well as to determine the temperature dependence of optical transitions associated with vacancies in \ce{MgO}, \ce{LiF}, and \ce{CsSnBr3}. 
Building on the recent work on defect line shapes \cite{LinOstWik25}, for each material, a \gls{nep} model was trained on \gls{dft} reference data encompassing defect configurations across the relevant charge states.
To explicitly represent the vacancy, the atoms neighboring the defect site were decorated with alternative atomic labels, enabling the potential to distinguish local environments associated with different charge states \cite{LinOstWik25}. 
Two distinct \gls{ti} paths were employed to accurately capture the temperature-induced changes in the \glspl{ctl}: the Frenkel–Ladd path \cite{reddy_review_frenkel_2021} and a direct integration path connecting \glspl{mlip} representing different charge states. 
Compared to previous finite temperature studies of defect free energies and \glspl{ctl}, our work provides a practical and general workflow for obtaining finite temperature \glspl{ctl} based on \glspl{mlip}, decomposes vibrational and band-edge (\gls{vbm}) contributions across three representative material classes, and demonstrates the finite temperature emergence of a neutral defect charge state in \ce{CsSnBr3} that is unstable at \qty{0}{\kelvin}.



The defect formation free energy for a defect $X$ in charge state $q$ is given by \autoref{eq:defect_formation_energy}.
\begin{align}
\begin{split}
    &G^f[X^q](T, \mu_e)
    = G_{\text{tot}}[X^q](T)
     - G_{\text{tot}}[\text{bulk}](T)
    \\
    &\quad\quad - \sum_i n_i \mu_i + q \left( E_{\text{VBM}}(T) + \mu_e \right)
     + E_{\text{corr}}(q).
\end{split}
    \label{eq:defect_formation_energy}
\end{align}
Here, $G_{\text{tot}}[X^q]$ and $G_{\text{tot}}[\text{bulk}]$ denote the Gibbs free energies of the defect-containing and pristine supercells, respectively.
The term $\sum_i n_i \mu_i$ accounts for the chemical potential contributions associated with adding or removing atoms of type $i$.
Since the \gls{ctl} does not depend on the chemical potentials, all $\mu_i$ are treated as constants throughout this study.
The term $q(E_{\text{VBM}} + \mu_e)$ represents the energy required to add or remove $q$ electrons referenced to the \gls{vbm}. 
The temperature dependence of the \gls{vbm} was obtained from \gls{md} snapshots via deep core level alignment, with further details given in the SI (Sec. 1.3).
$E_\text{corr}$ is the correction term for finite-size effects, which in the present work is charge dependent and evaluated using the \gls{fnv} correction scheme \cite{freysoldt_correction_2009, https://doi.org/10.1002/pssb.201046289}.
The values for ionic chemical potentials, temperature dependent \gls{vbm} shifts, and \gls{fnv} corrections are reported in the SI (Sec. 1.3).

\glspl{ctl} determine the thermodynamically favorable charge state of a defect at a given Fermi level.
The \gls{ctl} between charge states $q$ and $q'$, denoted as $\epsilon_{q/q'}$, is calculated from the difference in the defect formation free energies,
\begin{gather}
    \epsilon_{q/q'} = \frac{G^f[X^q;\mu_e=0]-G^f[X^{q'};\mu_e=0]}{q'-q}.
    \label{eq:ctl_position}
\end{gather}


\begin{figure*}
\centering
\includegraphics{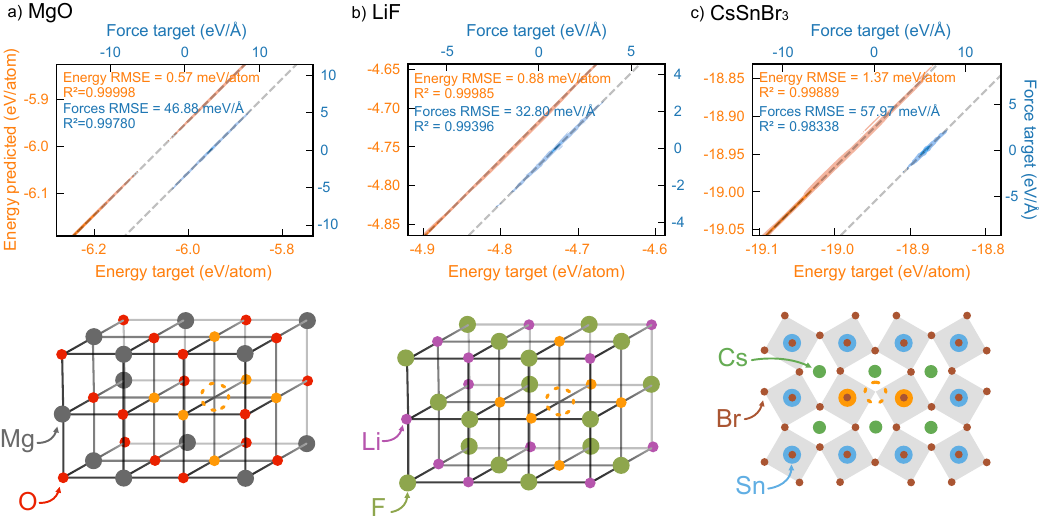}
\caption{
    \textbf{Machine-learned interatomic potential and illustrations of marked structures.}
    (a) The \ce{MgO} model performance. Six closest \ce{Mg} atoms to the oxygen vacancy are marked.
    (b) The \ce{LiF} model performance. Six closest \ce{Li} atoms to the fluorine vacancy are marked.
    (c) The \ce{CsSnBr3} model performance. Two closest \ce{Sn} atoms to the bromine vacancy are marked.
}
\label{fig:nep-model}
\end{figure*}

To obtain well-converged \gls{ti} results, we require an efficient way to perform \gls{md} simulations.
To this end, \gls{nep} models were trained for \ce{MgO}, \ce{LiF}, and \ce{CsSnBr3}. 
For \ce{MgO} and \ce{LiF}, the reference data were generated using the PBEsol exchange–correlation functional \cite{PhysRevLett.100.136406}, whereas for \ce{CsSnBr3} the r2SCAN functional \cite{furness_accurate_2020} was employed.
The training dataset comprised both pristine and defective structures. 
For \ce{MgO}, we included \ce{O} vacancies with charge states $0$ and $+2$; for \ce{LiF}, \ce{F} vacancies with charge states $-1$, $0$, and $+1$; and for \ce{CsSnBr3}, \ce{Br} vacancies with charge states $-1$, $0$, and $+1$.
In all training configurations, atoms neighboring a vacancy were labeled according to the charge state of that vacancy \cite{LinOstWik25}.
For \ce{MgO} and \ce{LiF}, the six \ce{Mg} or \ce{Li} atoms surrounding an \ce{O} or \ce{F} vacancy, respectively, were labeled, as illustrated in \autoref{fig:nep-model}. 
In \ce{CsSnBr3}, the two \ce{Sn} atoms closest to the defect were labeled. These labels encode the charge state of the associated vacancy.


To obtain free energy differences between charge states, including anharmonic contributions, we performed \gls{ti} directly between charge states and validated against Frenkel–Ladd \gls{ti}.

The direct path between charge states was implemented by duplicating the  \gls{mlip} and relabeling the atoms marked to represent a specific charge state. 
This produces two otherwise identical models that differ only in the assigned charge-state labeling of the defect environment.
We then perform \gls{ti} along a path between these two potentials, allowing us to compute the free-energy difference directly between the charge states.
The resulting free energy difference provided the formation energy difference between the charge states, from which the \glspl{ctl} were obtained.

The Frenkel–Ladd path \cite{reddy_review_frenkel_2021}, which connects the system described by the \gls{nep} model to a reference Einstein crystal with analytically known free energy, was used to calculate the free energy of both a pristine and defect cell.
The formation energy was then obtained from \autoref{eq:defect_formation_energy}.
This approach yielded the absolute free energy of each structure at the target temperature, enabling accurate determination of defect formation free energies. 
These formation energies for different charge states were subsequently used in \autoref{eq:ctl_position} to calculate the \glspl{ctl}.


\begin{figure*}
    \centering
    \includegraphics{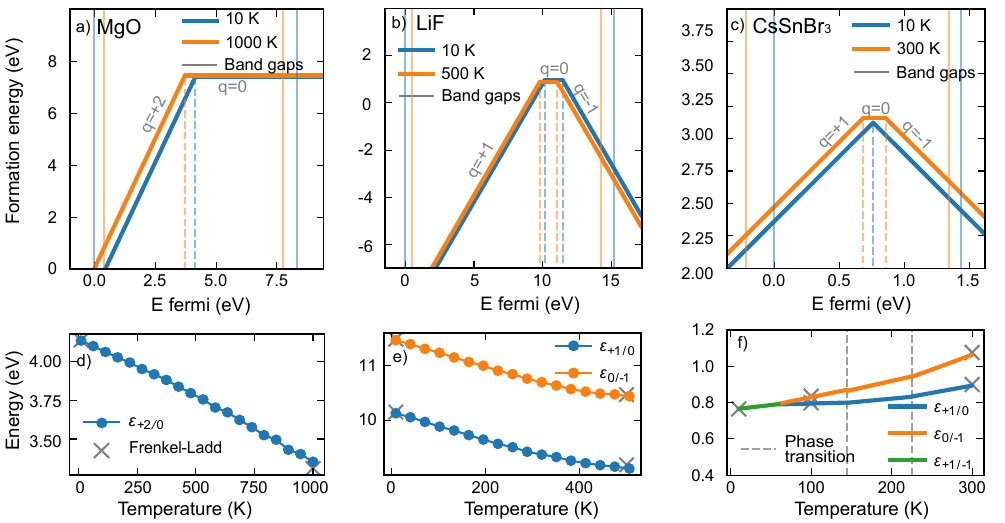}
    \caption{
    \textbf{Temperature-dependent \glspl{ctl}} for \ce{MgO}, \ce{LiF}, and \ce{CsSnBr3} with regard to \gls{vbm} position. Dashed lines represent the positions of the \glspl{ctl} and solid vertical lines indicate \gls{vbm} and \gls{cbm} positions, colored according to the temperature they represent.
    (a,d) The $\varepsilon_{+2/0}$ level in \ce{MgO} shifts by approximately \qty{-785}{\milli\electronvolt} over a temperature range of \qty{1000}{\kelvin}. (b,e) In \ce{LiF}, the $\varepsilon_{+1/0}$ and $\varepsilon_{0/-1}$ levels shift by about \qty{-994}{\milli\electronvolt} and \qty{-1035}{\milli\electronvolt}, respectively, over \qty{500}{\kelvin}.
    (c,f) The \glspl{ctl} exhibited a shift of roughly \qty{111}{\milli\electronvolt}, \qty{324}{\milli\electronvolt} and \qty{218}{\milli\electronvolt} over the \qty{300}{K} for the transitions between $+1/0$, $0/-1$ and $+1/-1$ respectively.
    Notably, a region where the neutral charge state is thermodynamically favorable is introduced around \qty{60}{\kelvin}. }
    \label{fig:ctl-ALL}
\end{figure*}

\begin{figure*}
\centering
\includegraphics{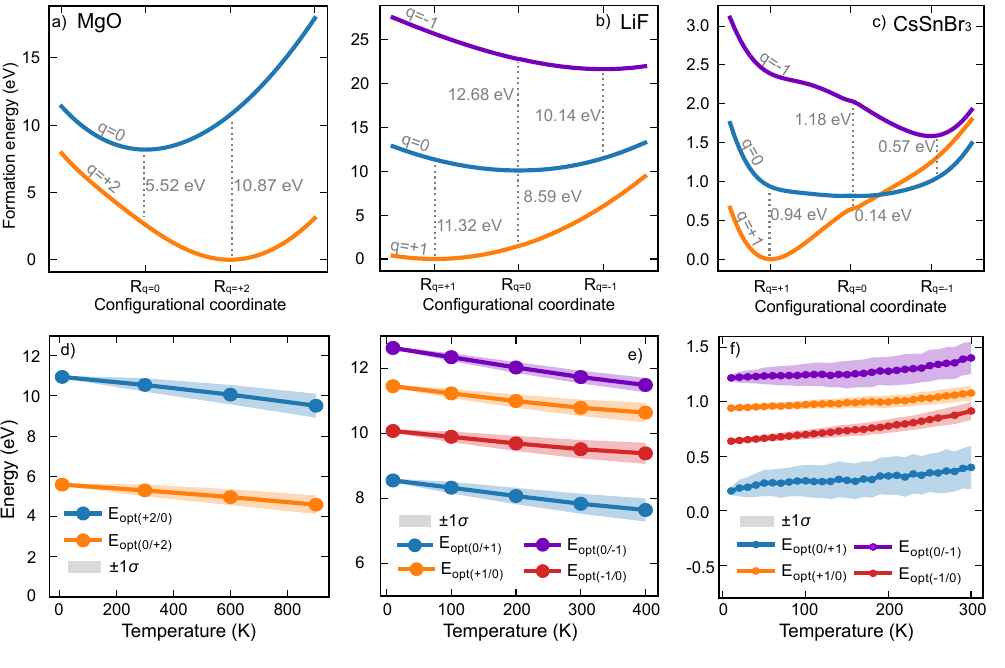}
\caption{
    \textbf{Configurational coordinate diagrams and temperature-dependent optical transitions} for \ce{MgO}, \ce{LiF}, and \ce{CsSnBr3}.
    (a--c) The energy difference at the relaxed geometry of each charge state defines the corresponding emission or absorption energy.
    The specific emission and absorption energies are shown as insets in the figures.
    (d--f) Emission and absorption energies during \gls{md} simulations at different temperatures.
    The shaded area represents the standard deviations of the sampled emission and absorption distributions.
}
\label{fig:optical-transitions}
\end{figure*}

The performance of the \gls{nep} models is shown in \autoref{fig:nep-model}.
All models achieve $R^2$ scores close to \num{1} and low \gls{rmse} scores.

Using \gls{ti} with our \gls{mlip}, we obtained the Gibbs free energy of the charge states and pristine cells by integrating along the Frenkel-Ladd path.
The defect formation energies were then evaluated using \autoref{eq:defect_formation_energy}, including the \gls{vbm} and correction terms.

These calculations enabled the construction of defect formation energy diagrams as a function of Fermi energy at various temperatures, as shown in \autoref{fig:ctl-ALL}a–c. 
In these plots, only the thermodynamically most stable charge state is shown at each Fermi energy. 
The positions of the \gls{vbm} and \gls{cbm} are marked by solid vertical lines, with colors corresponding to their respective temperatures.
The shifts of both \gls{vbm} and \gls{cbm} are reported in the SI (Sec. 1.3).
While we train our \gls{nep} models on data generated with semilocal functionals, we adjust the band edges in the formation energy plots for a more direct comparison with experimental data. 
It has been shown that \glspl{ctl} of localized defects are found at similar absolute energies at different levels of theory \cite{chen2013correspondence}. 
Therefore, we only adjust the positions of the \gls{vbm} and \gls{cbm} in the plots. 
The shifts are taken from comparisons between calculations performed at the PBEsol level of theory for MgO and LiF, and at the r2SCAN level for \ce{CsSnBr3}, against hybrid-functional calculations using the PBE0($\alpha$) functional.
In the case of \ce{CsSnBr3}, we additionally incorporate the shifts from including the spin-orbit coupling. 
For MgO and LiF, the values of the exact-exchange fraction ($\alpha = 0.34$ and $\alpha = 0.50$, respectively) are taken from Ref.~\cite{miceli2018nonempirical}, where they were determined using Koopmans’ theorem applied to the hydrogen probe and the H$^+$/0 level. 
For \ce{CsSnBr3} we use $\alpha = 0.26$, following Ref.~\citenum{bischoff2019nonempirical}, where the bromine vacancy was used as a probe. 

From the formation energy diagrams, the \glspl{ctl} at the temperatures where \gls{ti} was performed were extracted, shown as the $\times$-markers in \autoref{fig:ctl-ALL}d–f. 
Note that the shifts of \glspl{ctl} in these plots are reported relative to the temperature dependent \gls{vbm}.
Because the Frenkel–Ladd path requires extensive sampling to achieve convergence, the full temperature dependence of the \glspl{ctl} was obtained using the direct \gls{ti} path between charge states, also shown in \autoref{fig:ctl-ALL}d–f. 
The two \gls{ti} approaches yield closely agreeing results at all temperatures, while the direct path is substantially more computationally efficient. 
This enhanced efficiency is evident from the faster convergence shown in the SI (Fig. S7). 
This likely arises from the smaller energy differences along the \gls{ti} path between charge states compared with those encountered along the Frenkel–Ladd path.
Moreover, evaluating the free-energy difference between charge states via the Frenkel–Ladd route requires two separate \gls{ti} calculations, one for each charge state. 
In contrast, the direct charge-state path requires only a single \gls{ti} calculation to obtain the \gls{ctl}. 
Thus, the direct path both halves the minimum number of required \gls{ti} runs and converges substantially faster than the Frenkel–Ladd path.

Optical transition energies ($\text{E}_{\text{opt}(q'/q)}$ for transition from $q$ to $q'$) were calculated both at zero kelvin, using configurational coordinate diagrams (\autoref{fig:optical-transitions}a-c), and at finite temperatures, using \gls{md} simulations (\autoref{fig:optical-transitions}d-f).
These transitions are reported relative to the temperature dependent \gls{vbm} and move similarly to the associated \glspl{ctl}.
This is in agreement with previous results showing that the optical transition levels fluctuate similarly to the \gls{ctl} in \ce{CsPbBr3} \cite{CohEggRap19}.
The configuration coordinate diagram represents the \gls{pes} as the structure transforms between different defect geometries, where the energy differences between charge states at their minima correspond to emission or absorption processes. 
These diagrams were constructed by interpolating between the relaxed geometries of different charge states while the \gls{mlip} predicted the energy along the transformation path. 
These represent the zero-kelvin optical transitions, as they are derived from relaxed geometries.
At finite temperatures starting at \qty{10}{\kelvin}, \gls{md} trajectories of defect-containing cells were generated in charge state $q$, while the energies were simultaneously evaluated for charge state $q'$. 
The energy differences between these states correspond to the optical transition energies between states $q$ and $q'$, reflecting electronic excitations or absorption occurring on timescales too short for atomic relaxation. 
The temperature-dependent optical transition levels are shown in \autoref{fig:optical-transitions}d-f. 
The shaded regions represent the widths of the sampled emission and absorption distributions. 
For each temperature, the mean values and distribution widths were extracted from individual NPT ensemble simulation runs with durations ranging from \qty{0.2}{\nano\second} to \qty{5}{\nano\second}.


Oxygen vacancies in \ce{MgO} with charge states $+2$ and $0$ were investigated, resulting in one \gls{ctl}.
This \gls{ctl} exhibits a clear temperature dependence (\autoref{fig:ctl-ALL}a, d) that is driven by both vibrational entropy and a shift in \gls{vbm}.
The direct path between charge states yielded a \gls{ctl} shift of \qty{-0.785}{\milli\electronvolt\per\kelvin}.

Over the temperature range studied, the \gls{vbm} increases by approximately \qty{414}{\milli\electronvolt}, see Fig. S9 for details. 
This shift contributes to roughly half of the \gls{ctl} movement.
Excluding this shift from the total free energy change, the vibrational contribution from the path between different charge states averages \qty{-0.371}{\milli\electronvolt\per\kelvin}.

For \ce{MgO}, the optical transitions between the neutral and $+2$ charge states were considered.
From the configurational coordinate diagram (\autoref{fig:optical-transitions}a) these transitions appear as the energy difference between the states in the their respectively relaxed geometries.
We note that these zero kelvin vertical transitions are in reasonably good agreement with previous work done within the $G_0W_0$ approximation and the Bethe-Salpeter approach \cite{PhysRevLett.108.126404}.
Furthermore, the temperature dependence of these optical transitions and their fluctuation (\autoref{fig:optical-transitions}d) reveals a decrease in transition energies, which aligns well with the observed change in the \gls{ctl}.
In this figure, the shaded area indicates the standard deviation of the sampled optical transition energies, showing that these distributions widen with increased temperature.
This increase in distribution width is similar for both charge states.


For \ce{LiF}, fluorine vacancies with charge state $-1$, $0$ or $+1$ were investigated, leading to two \glspl{ctl} as all states have a region of stability.
Both \glspl{ctl} exhibit a shift with temperature (\autoref{fig:ctl-ALL}b, e).
The shifts in \glspl{ctl} from the direct integration are \qty{-1.988}{\milli\electronvolt\per\kelvin} and \qty{-2.070}{\milli\electronvolt\per\kelvin} for the $+1/0$ and $0/-1$ transitions, respectively.

The shift in \gls{vbm} over \qty{500}{\kelvin} was around \qty{605}{\milli \electronvolt} (see Fig. S11).
Excluding this from the \glspl{ctl} movement, the shifts become \qty{-0.778}{\milli\electronvolt\per\kelvin} and \qty{-0.860}{\milli\electronvolt\per\kelvin} for the $+1/0$ and $0/-1$ transitions, respectively.
This is a substantial shift, even when excluding the \gls{vbm} contribution, underscoring the importance of including vibrational entropy.

For \ce{LiF}, the optical transitions between the neutral and $+1/-1$ charge states were considered (\autoref{fig:optical-transitions}b, e).
As in the case of \ce{MgO}, the temperature dependence of these optical transitions and their fluctuations (\autoref{fig:optical-transitions}e) show a slight decrease in transition energies with increasing temperature, which aligns well with the observed change in the \gls{ctl}.
Furthermore, the distribution widths of different transitions are on the same scale.


For \ce{CsSnBr3}, bromine vacancies with charge state $-1$, $0$ or $+1$ were investigated.
In both the orthorhombic and tetragonal phases there are non-degenerate vacancy sites.
The difference in \gls{ctl} positions between these are about \qty{0.025}{\electronvolt} at zero kelvin and they all converge to the same values in the cubic phase as they become degenerate.
In this section, we report findings from the defect site that yields the lowest formation energy in the zero kelvin limit.
At absolute zero, only the negative and positive charge states are thermodynamically stable. 
However, at temperatures above approximately \qty{60}{\kelvin}, a region in which the neutral charge state becomes thermodynamically favored emerges (\autoref{fig:ctl-ALL}c).
Due to the soft dynamics of this material, independent \gls{ti} simulations fluctuated a lot in comparison to both \ce{LiF} and \ce{MgO}, requiring more sampling to reach convergence.
For this reason, linear fits of the movement of the \glspl{ctl} were performed.
Additionally, two phase transitions are observed across the temperature range considered, an orthorhombic to tetragonal transition at \qty{140}{\kelvin} and a tetragonal to cubic transition at \qty{225}{\kelvin}.
The linear fits were performed for each phase, showing minor changes in \gls{ctl} slope at these transitions (\autoref{fig:ctl-ALL}f).
Thus, structural phase transitions do not introduce discontinuous \gls{ctl} shifts for this defect.
The \glspl{ctl} exhibited shifts of \qty{0.370}{\milli\electronvolt\per\kelvin}, \qty{1.080}{\milli\electronvolt\per\kelvin}, and \qty{0.727}{\milli\electronvolt\per\kelvin} for the transitions between $+1/0$, $0/-1$, and $+1/-1$, respectively.

The \gls{vbm} shift (\qty{-214}{\milli\electronvolt} between 0 and \qty{300}{\kelvin}, see Fig. S13) accounts for a substantial part of the \gls{ctl} motion, but the sign and magnitude of the residual vibrational contribution differ between transitions, leading to the stabilization of the neutral state.
Excluding the \gls{vbm} shift, the \gls{ctl} moves by \qty{-0.343}{\milli\electronvolt\per\kelvin}, \qty{0.367}{\milli\electronvolt\per\kelvin}, and \qty{0.014}{\milli\electronvolt\per\kelvin} for the transitions between $+1/0$, $0/-1$, and $+1/-1$, respectively.
This highlights that the vibrational entropy is the effect stabilizing the neutral charge state at higher temperatures in \ce{CsSnBr3}.
These results show that, for \ce{CsSnBr3}, the \gls{ctl} moves slightly and has a substantial contribution from the change in the \gls{vbm}.
This small movement in the \glspl{ctl} due to vibrational entropy over this temperature range aligns well with previous results for halide perovskites \cite{mosquera2025dynamic}.
However, the results also show the importance of including finite temperature effects as the neutral charge state emerges as thermodynamically stable for some Fermi energies only above \qty{60}{\kelvin}.

For \ce{CsSnBr3}, the optical transitions between the neutral and $+1/-1$ charge states were considered (\autoref{fig:optical-transitions}c).
The optical transitions and their fluctuations (\autoref{fig:optical-transitions}f) exhibit an increase in transition energies.
This is consistent with the \glspl{ctl} of \ce{CsSnBr3} and also with previous studies showing that the optical transitions shifts together with \glspl{ctl} in \ce{CsPbBr3} \cite{CohEggRap19}.
Furthermore, the vertical energy distributions associated with transitions from the neutral to the charged states are substantially broader than for all other transitions. 
This is reflected by the flat energy landscape of the neutral charge state in comparison with both the negative and positive charge state (\autoref{fig:optical-transitions}c).
When calculating absorption or emission events from the neutral charge state during \gls{md}, the system naturally exhibits larger structural fluctuations, as the flat energy landscape allows it to move in and out of geometries corresponding to the energy minima of the negative and positive charge states.


The temperature dependence of the \glspl{ctl} across the investigated materials reveals distinct trends governed by the combination of vibrational entropy and \gls{vbm} shifts. 

For all materials investigated, substantial shifts in \glspl{ctl} were identified.
In \ce{MgO}, the \gls{ctl} movement is due to a near-equal shift in \gls{vbm} and vibrational entropy.
For \ce{LiF}, contributions from vibrational entropy alone accounts for significant movements in \gls{ctl} positions, which gets amplified when including the \gls{vbm} movement.
This emphasizes that lattice vibrations can drive significant \gls{ctl} shifts and shows that these effects are of importance when modeling defect energetics.
For \ce{CsSnBr3}, the temperature-induced movements of the \glspl{ctl} are moderate and influenced mainly by \gls{vbm} shifts.
Effects such as these have been measured experimentally, for instance in \ce{GaSe} where the difference between two \glspl{ctl}, from \ce{Ga} vacancies, increases from \qty{0.17}{\electronvolt} to \qty{0.38}{\electronvolt} over a range of \qty{290.5}{\kelvin} \cite{cryst15040372}.
This corresponds to a change of roughly \qty{0.722}{\milli\electronvolt\per\kelvin}, which is on the same order of magnitude as our results.
Temperature shifts in the optical transition energies (\autoref{fig:optical-transitions}d--f) have also been reported in previous experimental \cite{10.1063/1.1806545} and theoretical studies \cite{LinOstWik25}.

Furthermore, the most interesting observation related to the \glspl{ctl} in \ce{CsSnBr3} is the change that takes place at \qty{60}{\kelvin}, where the neutral charge state becomes stable at some Fermi energies.
The width of this stability range grows as temperature increases and at \qty{300}{\kelvin} the neutral charge state is the most stable one across a substantial part of the band gap.

Our results highlight two points: The inclusion of finite-temperature effects is of importance due to the potential temperature dependence of \glspl{ctl} and due to the possible introduction of new thermodynamically favorable charge states.


To conclude, we have quantified the temperature dependence of \glspl{ctl} and optical transition energies for vacancies in \ce{MgO}, \ce{LiF} and \ce{CsSnBr3} by employing \glspl{mlip}. 

The direct \gls{ti} path between charge states achieves high computational efficiency because the free energy differences along this path are small, in contrast to the much larger changes usually encountered in the Frenkel–Ladd path to an Einstein crystal. 
As a result, less sampling is required to obtain accurate estimates of the vibrational free-energy differences.

The results show that temperature can shift \glspl{ctl} by hundreds of meV, with the cause of these shifts differing between materials. 
In \ce{MgO}, \ce{LiF} and \ce{CsSnBr3} both vibrational and electronic (\gls{vbm}) contributions are relevant to  the movement of \glspl{ctl}.
Notably, in \ce{CsSnBr3} we find the emergence of a thermodynamically stable neutral charge state only above \qty{60}{\kelvin}.
This shows that even if \glspl{ctl} only vary slightly due to vibrational entropy contribution, this could have a significant impact by introducing entirely new thermodynamically stable charge states.
Finite temperature shifts in \glspl{ctl} and the emergence of new charge states as reported here are directly relevant for predicting doping limits and non-radiative recombination in wide-gap functional materials.

Together, these findings demonstrate that finite-temperature effects are essential for reliable predictions of defect energetics.
We show that the widely used static formalism of defect modeling fails to capture details that are important under operating conditions for these materials.

\section*{Data Availability}

All \gls{nep} models, as well as the \gls{dft} databases used to train these models, are available on Zenodo at \url{https://doi.org/10.5281/zenodo.17939378}

\section*{Acknowledgments}
We thank Sofia Cellini for useful comments. We acknowledge funding from the Swedish Strategic Research Foundation through a Future Research Leader programme (FFL21-0129), the Swedish Energy Agency (grant No. 45410-1), the Swedish Research Council (2018-06482, 2019-03993, and 2020-04935), the European Research Council (ERC Starting Grant No. 101162195), and the Knut and Alice Wallenberg Foundation (Nos.~2023.0032 and 2024.0042).
The computations were enabled by resources provided by the National Academic Infrastructure for Supercomputing in Sweden (NAISS) at C3SE, PDC, and NSC, partially funded by the Swedish Research Council through grant agreement no. 2022-06725.

\section*{Competing interests}

The authors declare that they have no competing interests.

\bibliography{references}

\end{document}